\documentclass[prl,twocolumn,reprint]{revtex4-1}
\usepackage{graphicx}
\usepackage{dcolumn}
\usepackage{bm}
\usepackage{color}
\usepackage{tabularx}
\usepackage{array}

\begin{document}

\title{Wavenumber-dependent Gilbert damping in metallic ferromagnets}

\author{Y. Li and W. E. Bailey}

\affiliation{Dept. of Applied Physics \& Applied Mathematics, Columbia University, New York NY 10027, USA}

\date{\today}

\begin{abstract}

A wavenumber-dependent dissipative term to magnetization dynamics, mirroring the conservative term associated with exchange, has been proposed recently for ferromagnetic metals. We present measurements of wavenumber- ($k$-) dependent Gilbert damping in three metallic ferromagnets, NiFe, Co, and CoFeB, using perpendicular spin wave resonance up to 26 GHz. In the thinnest films accessible, where classical eddy-current damping is negligible, size effects of Gilbert damping for the lowest and first excited modes support the existence of a $k^2$ term. The new term is clearly separable from interfacial damping typically attributed to spin pumping. Higher-order modes in thicker films do not show evidence of enhanced damping, attributed to a complicating role of conductivity and inhomogeneous broadening. Our extracted magnitude of the $k^2$ term, $\Delta\alpha^*_{kE}=\Delta\alpha_0^*+A_k^* k^2$ where $A_k^*$=0.08-0.1 nm$^2$ in the three materials, is an order of magnitude lower than that identified in prior experiments on patterned elements.

\end{abstract}

\maketitle

\indent The dynamical behavior of magnetization for ferromagnets (FMs) can be described by the Landau-Lifshitz-Gilbert (LLG) equation\cite{gilbertIEEE2004}:
\begin{equation}
\mathbf{\dot{m}}=-\mu_0|\gamma| \mathbf{m}\times\mathbf{H_{eff}}+\alpha \mathbf{m}\times\mathbf{\dot{m}}
\end{equation}
where $\mu_0$ is the vacuum permeability, $\mathbf{m}=\mathbf{M}/M_s$ is the reduced magnetization unit vector, $\mathbf{H_{eff}}$ is the effective magnetic field, $\gamma$ is the gyromagnetic ratio, and $\alpha$ is the Gilbert damping parameter. The LLG equation can be equivalently formulated, for small-angle motion, in terms of a single complex effective field along the equilibrium direction, as $\tilde{H}_{eff}$=$H_{eff}$-$i\alpha\omega/|\gamma|$; damping torque is included in the imaginary part of $\tilde{H}_{eff}$. \\
\indent For all novel spin-transport related terms to the LLG identified so far\cite{slonczewskiJMMM1996,bergerPRB1996,hirschPRL1999,sheNature2006,mironNmat2010,tserkovnyakPRL2002}, each real (conservative) effective field term is mirrored by an imaginary (dissipative) counterpart. In spin-transfer torque, there exist both conventional\cite{slonczewskiJMMM1996,bergerPRB1996} and field-like\cite{zhangPRL2002} terms in the dynamics. In spin-orbit torques (spin Hall\cite{hirschPRL1999} and Rashba\cite{mironNmat2010} effect) dampinglike and fieldlike components have been theoretically predicted\cite{stilesPRB2013} and most terms have been experimentally identified\cite{sheNature2006,mironNmat2010}. For pumped spin current\cite{tserkovnyakPRL2002}, theory predicts real and imaginary spin mixing conductances\cite{zwierzyckiPRB2005} $g^{\uparrow\downarrow}_r$ and $g^{\uparrow\downarrow}_i$ which introduce imaginary and real effective fields, respectively. \\
\indent It is well known that the exchange interaction, responsible for ferromagnetism, contributes a real effective field (fieldlike torque) quadratic in wavenumber $k$ for spin waves\cite{ckittel1951}. It is then natural to ask whether a corresponding imaginary effective field might exist, contributing a dampinglike torque to spin waves. Theoretically such an interaction has been predicted due to the \textit{intralayer} spin-current transport in a spin wave\cite{hankiewiczPRB2008,ForosPRB2008,tserkovnyakPRB2009,zhangPRL2009}, reflected as an additional term in Eq. (1):
\begin{equation}
\mathbf{\dot{m}}=\cdots - (|\gamma|\sigma_\perp /M_s) \mathbf{m}\times\nabla^2\mathbf{\dot{m}}
\end{equation}
where $\sigma_\perp$ is the transverse spin conductivity. This term represents a continuum analog of the well-established interlayer spin pumping effect\cite{tserkovnyakPRL2002,mizukamiPRB2002,ghoshAPL2011}. For spin wave resonance (SWR) with well-defined wavenumber $k$, Eq. (2) generates an additional Gilbert damping $\Delta\alpha(k)=(|\gamma|\sigma_\perp/M_s)k^2$. In this context, Gilbert damping refers to an intrinsic relaxation mechanism in which the field-swept resonance linewidth is proportional to frequency. Remarkably, the possible existence of such a term has not been addressed in prior SWR measurements. Previous studies of ferromagnetic resonance (FMR) linewidths of spin waves\cite{wigenPR1964,phillipsPL1964,gcbaileyJAP1970,fraitPRB1996} were typically operated at fixed frequency, not allowing separation of intrinsic (Gilbert) and extrinsic linewidths. Experiments have been carried out on thick FM films, susceptible to a large eddy current damping contribution\cite{ppincus1960}. Any wavenumber-dependent linewidth broadening in these systems has been attributed to eddy currents or inhomogeneous broadening, not intrinsic torques which appear in the LLG equation. \\
\indent In this Manuscript, we present a study of wavenumber-dependent Gilbert damping in the commonly applied ferromagnetic films Ni$_{79}$Fe$_{21}$ (Py), Co, and CoFeB. A broad range of film thicknesses (25-200 nm) has been studied in order to exclude eddy-current effects. We observe a thickness-dependent difference in the Gilbert damping for uniform and first excited spin wave modes which is explained well by the intralayer spin pumping model\cite{tserkovnyakPRB2009}. Corrections for interfacial damping, or conventional spin pumping, have been applied and are found to be small. The measurements show that the wavenumber-dependent damping, as identified in continuous films, is in reasonable agreement with the transverse spin relaxation lengths measured in Ref. \cite{ghoshPRL2012}, but an order of magnitude smaller than identified in experiments on sub-micron patterned Py elements\cite{nembachPRL2013}.\\
\indent Two different types of thin-film heterostructures were investigated in this study. Films were deposited by UHV sputtering with conditions given in Ref. \cite{ghoshPRL2012,yiJAP2013}. Multilayers with the structure Si/SiO$_2$(substrate)/Ta(5 nm)/Cu(5 nm)/\textbf{FM}($t_{FM}$)/Cu(5 nm)/Ta(5 nm), where \textbf{FM} = Py, Co and CoFeB and $t_{FM}$ = 25-200 nm, were designed to separate the effects of eddy-current damping and the intralayer damping mechanism proposed in Eq. (2). The minimum thickness investigated here is our detection threshold for the first SWR mode, 25 nm. A second type of heterostructure focused on much thinner Py films, with the structure Si/SiO$_2$(substrate)/Ta(5 nm)/Cu(5 nm)/Py($t_{Py}$)/Cu(5 nm)/\textbf{X}(5 nm), $t_{Py}$ = 3-30 nm. Here the cap layer \textbf{X} = Ta or SiO$_2$ was changed, for two series of this type, in order to isolate the effect of interfacial damping (spin pumping) from Cu/Ta interfaces. \\
\indent To study the Gilbert damping behavior of finite-wavenumber spin waves in the samples, we have excited perpendicular standing spin wave resonance (PSSWR)\cite{seaveyPRL1958} using a coplanar waveguide from 3 to 26 GHz. The spin-wave mode dispersion is given by the Kittel equation $\omega(k)/|\gamma|=\mu_0\left(H_{res} - M_s +H_{ex}(k)\right)$; the effective field from exchange, $\mu_0H_{ex}(k)=(2 A_{ex}/M_s)k^2$ with $A_{ex}$ as the exchange stiffness, gives a precise measurement of the wavenumber excited ((Fig. 1 \textit{inset})). PSSWR modes are indexed by the number of nodes $p$, with $k= p \pi/t_{FM}$ in the limit of unpinned surface spins. The full-width half-maximum linewidth, $\Delta H_{1/2}$, is fitted using $\mu_0\Delta H_{1/2}(\omega)=\mu_0\Delta H_0 + 2\alpha\omega/|\gamma|$ to extract the Gilbert damping $\alpha$. For $p=1$ modes we fix $\mu_0\Delta H_0$ as the values extracted from the corresponding $p=0$ modes for ($t_{FM}\leq40$ nm), because frequency ranges are reduced due to large exchange fields. In unconstrained fits for films of this thickness, the inhomogeneous broadening $\mu_0 \Delta H_0$ of the $p=1$ modes does not exhibit a discernible trend with $1/t^2_{FM}$ (or $k^2$)\cite{phillipsPL1964,gcbaileyJAP1970,fraitPRB1996}, justifying this approximation\cite{supplemental}. \\
\indent To fit our data, we have solved Maxwell's equations and the LLG equation (Eq. 1), including novel torques such as those given in Eq. (2), according to the method of Rado\cite{radoPR1955}. The model (designated 'EM+LLG') is described in the Supplemental Information. Values calculated using the EM+LLG model are shown with curves in Fig. 1 and dashed lines in Fig. 4. Comparison with such a model has been necessary since in our first type of sample series, $t_{FM}$ = 25-200 nm, eddy-current damping is negligible for thinner films (25 nm), the $A_k k^2$ contribution is negligible for thicker films (200 nm), but the two effects coexist for the intermediate region. \\
\indent In Fig. 1(a-c) we compare the measured Gilbert damping for the uniform ($p=0$, $\alpha_u$) and first excited ($p=1$, $\alpha_s$) spin wave modes. The dominant thickness-dependent contribution to Gilbert damping of the uniform modes of Py, Co, and CoFeB is clearly due to eddy currents which are quadratic in thickness. Note that eddy-current damping is negligible for the thinnest films investigated (25 nm), but quite significant for the thickest films (200 nm). This term sums with the bulk Gilbert damping $\alpha_0$\cite{scheckAPL2006}. The simulation of $\alpha_u$, shown by black curves in Fig. 1, matches closely with the analytical expression for bulk and eddy-current damping only\cite{JirsaPSS1982} of $\alpha_u = \alpha_{u0} + \alpha_{E0}$, where $\alpha_{E0} = {{\mu_0^2\gamma M_s t_{FM}^2} / {12 \rho_c}}$ denotes the eddy-current damping for uniform modes. Fittings of $\alpha_u$ yield resistivities $\rho_c=16.7$, 26.4 and 36.4 $\mu\Omega\cdot$cm for Py, Co and CoFeB, respectively.\\
\begin{figure}[htb]
\centering
\includegraphics[width = 3.0 in] {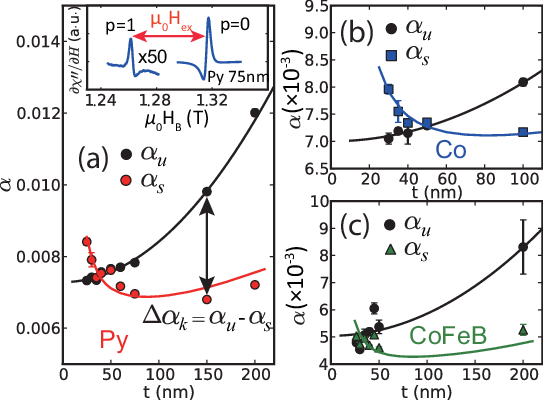}
\caption{Thickness dependence of $\alpha_u$ and $\alpha_s$ for (a) Py, (b) Co and (c) CoFeB thin films. Curves are calculated from a combined solution of Maxwell's equations and the LLG (EM+LLG). For $\alpha_u$ the values of $\mu_0M_s$, $\alpha$ (Table I), effective spin mixing conductance (Supplemental Information Section C) g-factor (2.12 for Py and CoFeB and 2.15 for Co) and $\rho_c$ (from analytical fitting) are used. For $\alpha_s$ the values of $\Delta\alpha_{kE}^*$ and $\Delta\alpha_{k0}^*$ (Table I) are also included in the simulation. \textit{Inset:} 10 GHz FMR spectra of $p=0$ and $p=1$ modes in Py 75 nm film.}
\end{figure}
\begin{figure}[htb]
\includegraphics[width = 3.0 in] {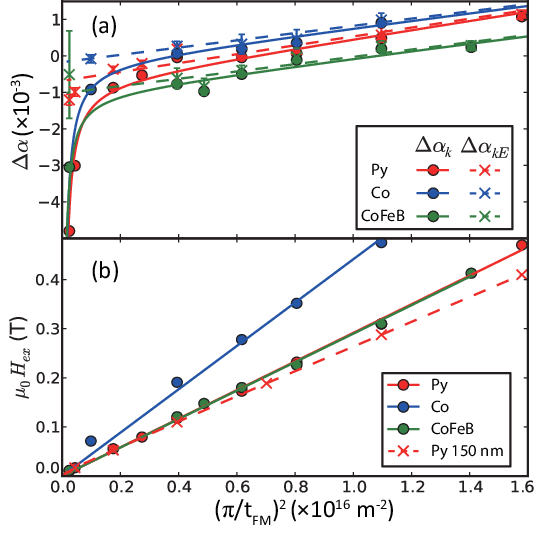}
\caption{Imaginary (damping, a) and real (exchange, b) effective fields as a function of $k^2$ for Py, Co and CoFeB. (a) Additional SWR damping $\Delta\alpha_k$ (circle) and  eddy-current corrected value $\Delta\alpha_{kE}$ (cross) as a function of $(\pi/t_{FM})^2$. Solid lines are guides to eye and dashed lines are fits to Eq. (3). (b) Exchange field $\mu_0H_{ex}$ as a function of $(\pi/t_{FM})^2$ ($(p\pi/t_{FM})^2$, $p$=0-6, for Py 150 nm). Lines are fits to $\mu_0H_{ex}=(2A/M_s) k^2$. }
\end{figure}
\indent Unlike the uniform-mode damping, the 1st SWR mode damping $\alpha_s$ is found to exhibit a minimum as a function of thickness. For decreasing thicknesses below 75 nm, $\alpha_s$ is increased. This behavior indicates an additional source of Gilbert damping for the 1st SWR modes. In CoFeB the increased $\alpha_s$ is less visible in Fig. 1(c) due to fluctuations in damping for samples of different thickness, but is evident in the difference, $\alpha_s-\alpha_u$, plotted in Fig. 2.\\
\indent In order to isolate this new damping mechanism, we plot in Fig. 2 the increased damping for the 1st SWR mode, $\Delta\alpha_k$=$\alpha_s-\alpha_u$, side-by-side with exchange field $\mu_0H_{ex}$ as a function of $(\pi/t_{FM})^2$ taken as the wavenumber $k^2$. When $\pi/t_{FM}$ is large, a linear $k^2$ dependence of $\Delta\alpha_{k}$ in all three ferromagnets mirrors the linear dependence of $\mu_0H_{ex}$ on $k^2$. This parallel behavior reflects the wavenumber-dependent imaginary and real effective fields acting on magnetization, respectively. To quantify the quadratic wavenumber term in $\Delta\alpha_k$, we also show the eddy-current-corrected values $\Delta\alpha_{kE}$=$\Delta\alpha_k-\Delta\alpha_E$ in Fig. 2(a). Here $\Delta\alpha_E=\alpha_{E1}-\alpha_{E0}$ denotes the difference in eddy current damping between $p=1$ and $p=0$ modes according to the theory of Ref. \cite{JirsaPSS1982}, for weak surface pinning, where $\alpha_{E1}\approx0.23\alpha_{E0}$ (See Supplemental Information for more details). We then fit this eddy-current-corrected value to a linearization of Eq. (2), as:
\begin{equation}
\Delta\alpha_{kE} = \Delta\alpha_{k0} + A_k k^2
\end{equation}
with $A_k=|\gamma|\sigma_\perp/M_s$ and $\Delta\alpha_{k0}$ a constant offset. The values of $A_k$ estimated this way are $0.128\pm0.022$ nm$^2$, $0.100\pm0.011$ nm$^2$ and $0.100\pm0.018$ nm$^2$ for Py, Co and CoFeB. \\
\indent Recently, Kapelrud et al.\cite{brataasPRL2013} have predicted that interface-localized (e.g. spin-pumping) damping terms will also be increased in SWR, with interfacial terms for $p\geq 1$ modes a factor of two greater than those for the $p=0$ mode. Using the second series of thinner Py films, we have applied corrections for the interfacial term to our data, and find that these effects introduce only a minor ($\sim$20\%) correction to the estimate of $A_k$. The $p=0$ mode damping associated with the Cu/Ta interface has been measured from the increase in damping upon replacement of SiO$_2$ with Ta at the top surface (Fig. 3, \textit{inset}). Here Cu/SiO$_2$ is taken as a reference with zero interfacial damping; insulating layers have been shown to have no spin pumping contribution\cite{mosendzAPL2009}. We find the damping enhancement to be inversely proportional to $t_{FM}$, indicating an interfacial damping term quantified as spin pumping into Ta\cite{tserkovnyakPRL2002} with $\Delta\alpha_{sp} = {{\gamma\hbar (g^{\uparrow\downarrow}/S)}/{4\pi M_s}t_{FM}}$. Using the values in Table I yields the effective spin mixing conductance as $g^{\uparrow\downarrow}_{Py/Cu/Ta}/S$=2.5 nm$^{-2}$, roughly a factor of three smaller than that contributed by Cu/Pt interfaces\cite{ghoshAPL2011}. \\
\indent Using the fitted $g^{\uparrow\downarrow}_{FM/Cu/Ta}/S$, we calculate and correct for the additional spin pumping contribution to damping of the $p=1$ mode, $2\Delta\alpha_{sp}$ (from top and bottom interfaces). The corrected values for the 1st SWR damping enhancement, $\Delta\alpha_{kE}^* = \Delta\alpha_{kE}-2\Delta\alpha_{sp}$, are plotted for Py(25-200nm) in Fig. 3. These corrections do not change the result significantly. We fit the $k^2$ dependence of $\Delta\alpha_{kE}^*$ to Eq. (3) to extract the corrected values $A_k^*$ and $\Delta\alpha_{k0}^*$. The fitted value, $A_k^*=0.105\pm0.021$ nm$^2$ for Py, is slightly smaller than the uncorrected value $A_k$. Other extracted interfacial-corrected values $A^*_k$ are listed in Table I. Note that the correction of wavenumber by finite surface anisotropy will only introduce a small correction of $A_k$ and $A_k^*$ within errorbars. We also show the EM+LLG numerical simulation results for the uniform modes and the first SWR modes in Fig. 1 (solid curves). Those curves coincide with the analytical expressions of eddy-current damping plus $k^2$ damping (not shown) and fit the experimental data points nicely. \\
\begin{figure}[htb]
\centering
\includegraphics[width = 3.0 in] {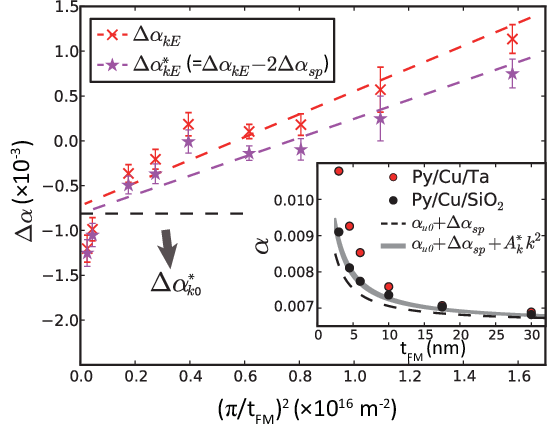}
\caption{Interfacial damping correction for Py. Main panel: $\Delta\alpha_{kE}$ and $\Delta\alpha^*_{kE}$ as a function of $(\pi/t_{FM})^2$. Dashed lines are fits to $k^2$-dependent equation as Eq. (3); $\Delta\alpha_{k0}^*$ are extracted from $\Delta\alpha^*_{kE}$ fits. {\it Inset:} size effect of uniform-modes Gilbert damping in Py/Cu/Ta and Py/Cu/SiO$_2$ samples (circles). The dashed curve is the theoretical reproduction of Py/Cu/SiO$_2$ using $\alpha_{u0}+\Delta\alpha_{sp}(t_{FM})$. The shadow is the same reproduction using $\alpha_{u0}+\Delta\alpha_{sp}(t_{FM})+A_k^*k^2$ where the error of shadow is from $A_k^*$. Here $k$ is determined by $A_{ex} k^2 = 2 K_s / t_{FM}$.}
\end{figure}
\indent The negative offsets $\Delta\alpha_{k0}^*$ between uniform modes and spin wave modes for Py and CoFeB are attributed to resistivitylike intrinsic damping\cite{gilmorePRL2007}: because $\mathbf{\dot{m}}$ is averaged through the whole film for uniform modes and maximized at the interfaces for unpinned boundary condition, the SWR mode experiences a lower resistivity near low-resistivity Cu and thus a reduced value of damping. For Co a transition state between resistivitylike and conductivitylike mechanisms\cite{smbhagatPRB1974} corresponds to negligible $\Delta\alpha_{k0}^*$ as observed in this work.\\
\indent In addition to the thickness-dependent comparison of $p=0$ and $p=1$ modes, we have also measured Gilbert damping for a series of higher-order modes in a thick Py (150 nm) film. Eddy-current damping ($\alpha_E\sim0.003$) is the dominant mode-dependent contribution in this film. The wavenumber $k$ for the mode $p=6$ is roughly equal to that for the first SWR, $p=1$, in the 25 nm film. Resonance positions are plotted with the dashed lines in Fig. 2(b), as a function of $k$, and are in good agreement with those found from the $p=1$ data. In Fig. 4 we plot the mode-related Gilbert damping $\alpha_p$ up to $p=6$, which gradually decreases as $p$ increases. We have again conducted full numerical simulations using the EM+LLG method with ($A_k^* = 0.105$ nm$^2$) or without ($A_k^*=0$) the intralayer spin pumping term, shown in red and black crosses, respectively. Neither scenario fits the data closely; an increase at $p=3$ is closer to the model including the $k^2$ mechanism, but experimental $\alpha$ at $p=6$ falls well below either calculation. \\
\indent We believe there are two possibilities why the $\alpha \propto p^2$ damping term is not evident in this configuration. First, the effective exchange field increases with $p$, resulting in a weaker (perpendicular) resonance field at the same frequency. When the perpendicular biasing field at resonance is close to the saturation field, the spins near the boundary are not fully saturated, which might produce an inhomogeneous linewidth broadening at lower frequencies and mask small Gilbert contributions from wavenumber effect. From the data in Fig. 4 \textit{inset} the high-$p$ SWR modes is more affected by this inhomogeneous broadening and complicate the extraction of $k^2$ damping. Second, high-$p$ modes in thick films are close to the anomalous conductivity regime, $k\lambda_M \sim 1$, where $\lambda_M$ is the electronic mean free path. The Rado-type model such as that applied in Fig. 4 is no longer valid in this limit\cite{radoJAP1958}, beyond which Gilbert damping has been shown to decrease significantly in Ni and Co\cite{korenmanPRB1972}. Based on published $\rho\lambda_M$ products for Py\cite{gurneyPRL1993} and our experimental value of $\rho_c=16.7$ $\mu\Omega\cdot$cm, we find $\lambda_M\sim 8$ nm and $k\lambda_M \sim 1$ for the $p=6$ mode in Py 150 nm. For the 1st SWR mode in Py 25 nm, on the other hand, eddy currents are negligible and the anomalous behavior is likely suppressed due to surface scattering, which reduces $\lambda_M$. \\
\indent An important conclusion of our work is that the intralayer spin pumping, as measured classically through PSSWR, is indeed present but more than 10 times smaller than estimated in single nanoscale ellipses\cite{nembachPRL2013}. The advantages of the PSSWR measurements presented in this manuscript are that the one-dimensional mode profile is well-defined, two-magnon effects are reduced, if not absent\cite{mcmichaelJAP1998}, and there are no lithographic edges to complicate the analysis. The lower estimates of $A_k^*$ from PSSWR are sensible, based on physical parameters of Py, Co, and CoFeB. The polarization of continuum-pumped spins in a nearly uniformly magnetized film, like that of pumped spin current in a parallel-magnetized F/N/F structure, is transverse to the magnetization\cite{tserkovnyakPRB2009}. From the measured transverse spin conductance $\sigma_\perp$ we extract that the relaxation lengths of pumping intralayer spin current are 0.8-1.9 nm for the three ferromagnets\cite{supplemental}, in good agreement with the small transverse spin coherence lengths found in these same ferromagnetic metals\cite{zhangPRL2004,ghoshPRL2012}.\\
\begin{figure}[htb]
\centering
\includegraphics[width = 3.0 in] {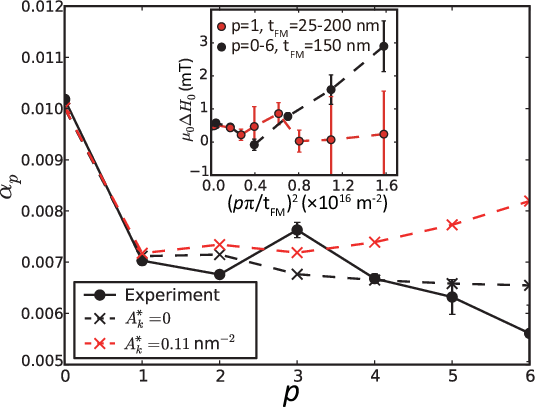}
\caption{Mode-dependent damping $\alpha_p$ for Py(150nm), $0\leq p\leq 6$.  Crosses are EM+LLG calculated values with and without the wavenumber-dependent damping term. {\it Inset}: Inhomogeneous broadening $\Delta H_0$ vs $0\leq p\leq 6$, 150nm film. Larger, k-dependent values are evident, compared with those in the thickness series ($t_{FM}$=25-200 nm).}
\end{figure}
\indent Finally, we show that the magnitude of the intralayer spin pumping identified here is consistent with the damping size effect {\it not} attributable to interlayer spin pumping, in layers without obvious spin sinks. For the $p=0$ mode, a small but finite wavenumber is set by the surface anisotropy through\cite{soohooPR1963,JirsaPSS1982} $A_{ex} k^2 = 2 K_s / t_{FM}$. The damping enhancement due to intralayer spin pumping will, like the interlayer spin pumping, be inverse in thickness, leading to an 'interfacial' term as $\alpha = 2 K_s (A^*_{k}/A_{ex}) t_{FM}^{-1}$. This contribution is indicated by the grey shadow in Fig. 3 \textit{inset} and provides a good account of the additional size effect in the SiO$_2$-capped film. Here we use $K_s$=0.11 mJ/m$^2$ extracted by fitting the thickness-dependent magnetization to $\mu_0M_{eff}=\mu_0M_s-4K_s/M_s t_{FM}$. While alternate contributions to the observed damping size effect for the SiO$_2$-capped film cannot be ruled out, the data in Fig. 3 \textit{inset} place an upper bound on $A_k^*$. \\
\indent In summary, we have identified a wavenumber-dependent, Gilbert-type damping contribution to spin waves in nearly uniformly magnetized, continuous films of the metallic ferromagnets Py, Co and CoFeB using classical spin wave resonance. The term varies quadratically with wavenumber, $\Delta\alpha \sim A_k^* k^2$, with the magnitude, $A_k^* \sim 0.08$-$0.10$ nm$^2$, amounting to $\sim$20\% of the bulk damping in the first excited mode of a 25 nm film of Py or Co, roughly an order of magnitude smaller than previously identified in patterned elements. The measurements quantify this texture-related contribution to magnetization dynamics in the limit of nearly homogeneous magnetization. \\

\begin{table}[ht]
\centering
\begin{tabular}{>{\centering\arraybackslash}m{0.35in} >{\centering\arraybackslash}m{0.5in} >{\centering\arraybackslash}m{0.5in} >{\centering\arraybackslash}m{0.6in} >{\centering\arraybackslash}m{0.6in} >{\centering\arraybackslash}m{0.45in}}
\hline
& $\mu_0M_{s}$(T) & $\alpha_0$ & $A_{ex}$(J/m) & $A_k^*$(nm$^2$) & $\Delta\alpha_0^*$ \\
\hline\hline
Py    & 1.00 & 0.0073 & 1.2$\times10^{-11}$ & $0.11\pm0.02$ & -0.0008  \\
Co    & 1.47 & 0.0070 & 3.1$\times10^{-11}$ & $0.08\pm0.01$ & -0.0002  \\
CoFeB & 1.53 & 0.0051 & 1.8$\times10^{-11}$ & $0.09\pm0.02$ & -0.0011  \\
\hline
\end{tabular}
\caption{Fit parameters extracted from resonance fields and linewidths of uniform and 1st SWR modes. Values of $A_k^*$ and $\Delta\alpha_0^*$ for Co and CoFeB are calculated using the spin mixing conductances measured in FM/Cu/Pt\cite{ghoshAPL2011}. See the Supplemental Material for details.}
\label{table2}
\end{table}

\bibliographystyle{apsrev4-1}

\end{document}